\documentclass[twocolumn,showpacs,preprintnumbers,amsmath,amssymb]{revtex4}
\usepackage[dvips]{graphicx}

\def\lsim{\buildrel {\textstyle <}\over {_\sim}}

\begin{document}
\title{Numerical evidence of the spin-chirality decoupling in the three-dimensional Heisenberg spin glass
}
\author{Dao Xuan Viet and Hikaru Kawamura}
\affiliation{Department of Earth and Space Science, Faculty of Science,
Osaka University, Toyonaka 560-0043,
Japan}
\date{\today}
\begin{abstract}
Ordering of the three-dimensional Heisenberg spin glass with Gaussian coupling is studied by extensive Monte Carlo simulations. 
The model undergoes successive chiral-glass and spin-glass transitions at nonzero temperatures $ T_{CG} > T_{SG} > 0$, exhibiting the spin-chirality decoupling.
\end{abstract}
\maketitle

 The issue of the spin-glass (SG) ordering has been studied quite extensively for years, and continued to give an impact on surrounding areas. Meanwhile, the original problem of the magnetic ordering of typical spin-glass magnets {\it e.g.\/}, canonical SG, still remains to be elusive \cite{review}. As magnetic interactions in  many SG materials are nearly isotropic, it is important to elucidate the ordering properties of the three-dimensional (3D) isotropic Heisenberg SG.  Although earlier numerical studies suggested that the 3D Heisenberg SG exhibited only a $T=0$ transition \cite{Olive,Matsubara91}, one of the present authors (H.K.) suggested that the model might exhibit a finite-temperature transition {\it in its chiral sector\/} \cite{Kawamura92}. Chirality is a multispin variable representing the handedness of the noncollinear or noncoplanar structures induced by frustration. It has subsequently been suggested that, in the ordering of the 3D Heisenberg SG, the chirality was ``decoupled'' from the spin, the chiral-glass (CG) order taking place at a temperature higher than the SG order, $T_{CG} > T_{SG}$ \cite{HukuKawa00,HukuKawa05,Kawamura07}. Based on such a spin-chirality decoupling picture of the 3D isotropic Heisenberg SG, a chirality scenario of experimental SG transition was proposed \cite{Kawamura92,Kawamura07}: According to this scenario, the chirality is a hidden order parameter of SG transition. Real SG transition of weakly anisotropic SG magnets is then a ``disguised'' CG transition, where the chirality is mixed into the spin sector via a weak random magnetic anisotropy.

 Although consensus now seems to appear in recent numerical studies that the 3D Heisenberg SG indeed exhibits a finite-temperature transition \cite{Kawamura92,HukuKawa00,HukuKawa05,Kawamura07,Matsubara00,LeeYoung03,Campos06,LeeYoung07}, the nature of the transition, especially whether the model really exhibits the spin-chirality decoupling, is still under hot debate. Obviously, it is crucially important to clarify the ordering of the 3D Heisenberg SG model. 

 The present situation, however, is not completely satisfactory.  Mentioning some of the recent numerical works: By simulating the model of modest lattice sizes $L\leq 20$ ($L$ being the linear dimension) but with rather small number of samples of $N_s=32$ (for their largest $L$), Hukushima and Kawamura presented support for the spin-chirality decoupling \cite{HukuKawa05}. By contrast, Lee and Young claimed on the basis of their data of the correlation-length ratios $\xi/L$ that the spin and the chirality order at a common temperature, thus no spin-chirality decoupling \cite{LeeYoung03,LeeYoung07}. Their data, however, suffers from either small lattice sizes of only $L\leq 12$ \cite{LeeYoung03} or small number of samples of $N_s=56$ \cite{LeeYoung07}. Campos {\it et al\/} simulated the same model to much larger lattices $L=32$ with larger number of samples $N_s=1,000$, but no data below $T_g$ \cite{Campos06}. Campos {\it et al\/} claimed that the chiral and spin sectors undergo simultaneously a Kosterlitz-Thouless (KT) transition with massive logarithmic corrections. This interpretation, however, was criticized in Ref.\cite{Campbell}. 

 Under such circumstances, we perform here a large-scale Monte Carlo simulation of the 3D Heisenberg SG in order to shed further light on the nature of its spin and chirality ordering. We exceed the previous simulations by simulating the system as large as $L=32$ to temperatures considerably lower than $T_g$ for large number of samples of order $N_s\simeq 10^3$. Note that none of the previous simulations satisfied all these criteria simultaneously. More importantly, we calculate several independent physical quantities including the correlation-length ratios, the Binder ratios and the glass order parameters, trying to draw consistent picture from these independent quantities, whereas Refs.\cite{LeeYoung03,Campos06,LeeYoung07} concentrated almost exclusively on the correlation-length ratio.  Our simulation then enabled us to conclude that the SG transition occurs at a nonzero temperature which is located about 15\%  below the CG transition temperature. Thus, the 3D Heisenberg SG certainly exhibits the spin-chirality decoupling.

 The model is the isotropic classical Heisenberg model on a 3D simple cubic lattice with the nearest-neighbor Gaussian coupling. The Hamiltonian is given by
\begin{equation}
{\cal H}=-\sum_{<ij>}J_{ij}\vec{S}_i\cdot \vec{S}_j\ \ ,
\label{eqn:hamil}
\end{equation}
where $\vec{S}_i=(S_i^x,S_i^y,S_i^z)$ is a three-component unit vector at the $i$-the site, and the $<ij>$ sum is taken over nearest-neighbor pairs. The coupling $J_{ij}$ are random Gaussian variables with zero mean and the variance $J^2$. 

 The local chirality at the $i$-th site and in the $\mu$-th direction $\chi_{i\mu}$ may be defined for three neighboring Heisenberg spins by the scalar
\begin{equation}
\chi_{i\mu}=
\vec{S}_{i+{\hat{e}}_{\mu}}\cdot
(\vec{S}_i\times\vec{S}_{i-{\hat{e}}_{\mu}})\ \ ,
\end{equation}
where ${\hat{e}}_{\mu}\ (\mu=x,y,z)$ denotes a unit vector along the $\mu$-th axis. There are in total $3N$ local chiral variables.

 The lattice contains $N=L^{3}$ sites with $L=6$, 8, 12, 16, 24, 32 with periodic boundary conditions. Sample average is taken over 2000 ($L=6,8,12$), 1000 ($L=16, 24$) and 800 ($L=32$) bond realizations. To facilitate efficient thermalization, we employ the single-spin-flip heat-bath and over-relaxation method \cite{Campos06}, combined with the temperature-exchange technique. Over-relaxation sweep is repeated $L$ times per every heat-bath sweep.

 Care is taken to make sure that the system is fully equilibrated. Equilibration is checked by the following procedures. First, we monitor the system to travel back and forth many times along the temperature axis during the temperature-exchange process (typically more than 10 times) between the maximum and minimum temperatures, while we also check that the relaxation due to the single-spin-flip updating is fast enough at the highest temperature. This guarantees that different parts of the phase space are sampled in each ``cycle'' of the temperature-exchange run. Second, we follow Ref.\cite{Katzgraber} and check the equality expected to hold for the model with Gaussian couping. Third, we check the stability of the results against at least three times longer runs for a subset of samples. Fourth, we compare the data of the correlation-length ratios with the recent data by other authors in the temperature range where common data are available \cite{Campos06,LeeYoung07}. Error bars are estimated by the sample-to-sample statistical fluctuations.

 We run two independent systems (1) and (2), and calculate a $k$-dependent overlap. For the chirality, the $k$-dependent chiral overlap $q_\chi(\vec k)$ is defined by the scalar,
\begin{equation}
q_\chi(\vec k) =
\frac{1}{3N}\sum_{i=1}^N\sum_{\mu=x,y,z}
\chi_{i\mu}^{(1)}\chi_{i\mu}^{(2)}e^{i\vec k\cdot \vec r_i},
\end{equation}
whereas, for the spin, it is defined by the {\it tensor\/} $q_{\alpha\beta}(\vec k)$ between the $\alpha$ and $\beta$ components of the Heisenberg spin,
\begin{equation}
q_{\alpha\beta}(\vec k) = 
\frac{1}{N}\sum_{i=1}^N S_{i\alpha}^{(1)}S_{i\beta}^{(2)}e^{i\vec k\cdot \vec r_i},
\ \ \ (\alpha,\beta=x,y,z).
\end{equation}
The CG and SG order parameters are defined by the second moment of the $k=0$ component of the overlap,
\begin{equation}
q_{CG}^{(2)}=[\langle q_{\chi}(\vec 0)^2\rangle],
\end{equation}
\begin{equation}
q_{SG}^{(2)} = [\langle q_{\rm s}(\vec 0)^2\rangle]\ \ ,
\ \ \ \ 
q_{\rm s}(\vec k)^2 = \sum_{\alpha,\beta=x,y,z} \left| q_{\alpha\beta}(\vec k) \right| ^2,
\end{equation}
where $\langle\cdots\rangle$ represents the thermal average and [$\cdots$] the average over the bond disorder. The chiral and spin Binder ratios are defined by
\begin{equation}
g_{CG}=
\frac{1}{2}
\left(3-\frac{[\langle q_{\chi}(\vec 0)^4\rangle]}
{[\langle q_{\chi}(\vec 0)^2\rangle]^2}\right),
\end{equation}
\begin{equation}
g_{SG} = \frac{1}{2}
\left(11 - 9\frac{[\langle q_{\rm s}(\vec 0)^4\rangle]}
{[\langle q_{\rm s}(\vec 0)^2\rangle]^2}\right).
\label{eqn:gs_def}
\end{equation}
Note that both $g_{CG}$ and $g_{SG}$ are normalized so that they vanish in the high-temperature phase for $L\rightarrow \infty$ and gives unity in the nondegenerate ground state as expected for the present Gaussian coupling.

 The finite-size correlation lengths are given by \cite{LeeYoung03}
\begin{equation}
\xi = 
\frac{1}{2\sin(k_\mathrm{m}/2)}
\sqrt{ \frac{ [\langle q(\vec 0)^2 \rangle] }
{[\langle q(\vec{k}_\mathrm{m})^2 \rangle] } -1 },
\end{equation}
for each case of the chirality and the spin, $\xi_{CG}$ and $\xi_{SG}$, where $\vec{k}_{\rm m}=(2\pi/L,0,0)$ with $k_{\textrm{m}}=|\vec k_{\textrm{m}}|$, and the $\mu$-direction in Eq.(2) is taken here being parallel with $\vec k$.

\begin{figure}[ht]
\begin{center}
\includegraphics[scale=0.9]{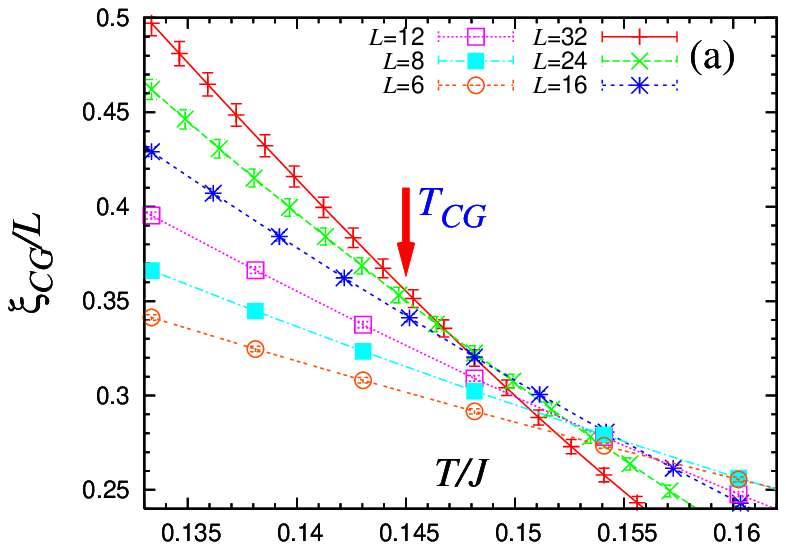}
\includegraphics[scale=0.9]{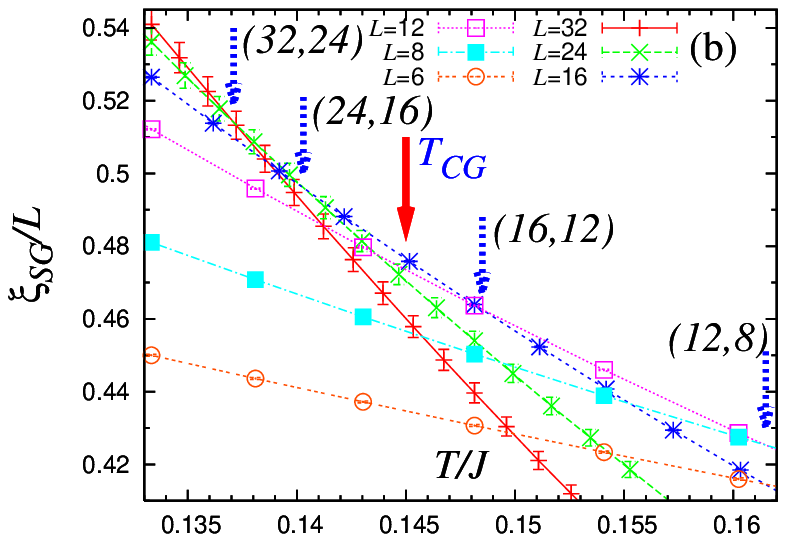}
\end{center}
\caption{
The temperature and size dependence of the correlation-length ratio for the chirality (a), and for the spin (b). The arrow indicates the bulk chiral-glass transition point.
}
\end{figure}
\begin{figure}[ht]
\begin{center}
\includegraphics[scale=0.9]{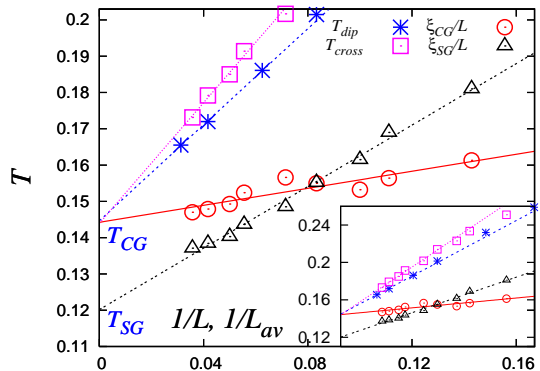}
\end{center}
\caption{
The (inverse) size dependence of the crossing temperatures of $\xi_{CG}/L$ and  $\xi_{SG}/L$, the dip temperature $T_{dip}$ and the crossing temperature $T_{cross}$ of $g_{CG}$. From the linear extrapolation of the data, the sin-glass and chiral-glass transition temperatures are estimated as $T_{CG}=0.145\pm 0.004$ and  $T_{SG}=0.120\pm 0.006$. The inset exhibits a wider range.
}
\end{figure}
\begin{figure}[ht]
\begin{center}
\includegraphics[scale=0.9]{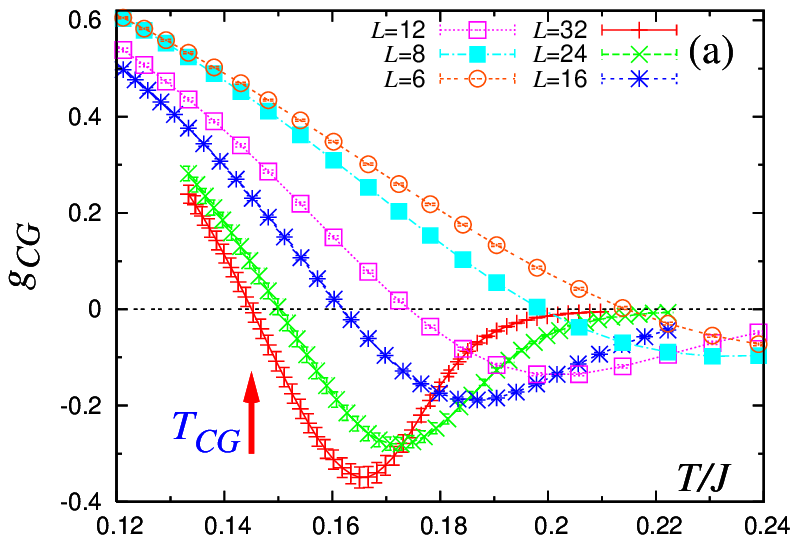}
\includegraphics[scale=0.9]{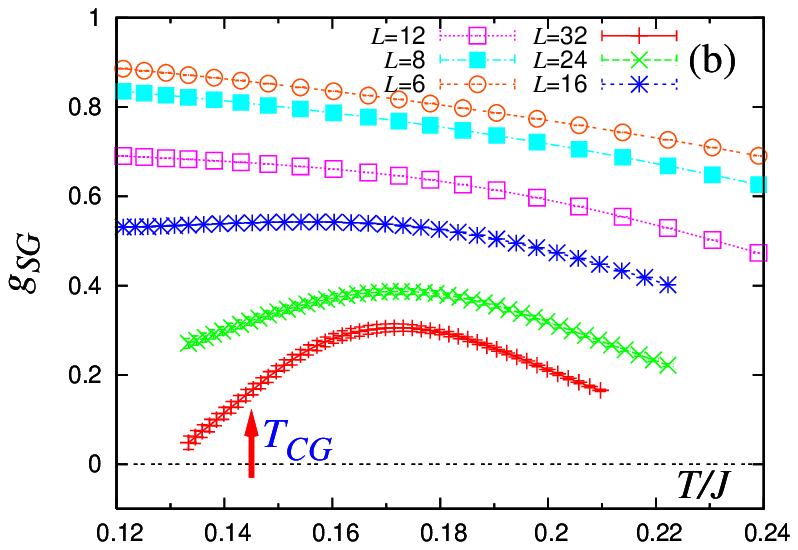}
\end{center}
\caption{
The temperature and size dependence of the Binder ratio for the chirality (a), and for the spin (b). The arrow indicates the bulk chiral-glass transition point.}
\end{figure}
 In Fig.1, we show the correlation-length ratios for the chirality  $\xi_{CG}/L$ (a), and for the spin  $\xi_{SG}/L$ (b). While the chiral $\xi_{CG}/L$ curves cross at temperatures which are only weakly $L$-dependent, the spin $\xi_{SG}/L$ curves cross at progressively lower temperatures as $L$ increases.  The $\xi/L$ data are compared with the data by other authors as follows: Our data for $\xi/L$ are in full agreement with those of Ref.\cite{Campos06} within statistical error bars over the narrow and relatively high-temperature range covered by their data. The data of Ref.\cite{LeeYoung07} for their largest $L$ (on which their claim for ``merging'' was based) are lower than our present ones and those of Ref.\cite{Campos06} by about 5 to 6 of our $\sigma$ units; this may be a purely statistical effect in view of the limited number of samples measured in Ref.\cite{LeeYoung07}.


 To estimate the bulk CG and SG transition temperatures quantitatively, we plot in Fig.2 the crossing temperature of $\xi_{CG}/L$ and $\xi_{SG}/L$ for pairs of successive $L$ values versus $1/L_{av}$, where $L_{av}$ is a mean of the two sizes. The data show an almost linear $1/L_{av}$-dependence. The chiral crossing temperature exhibits a weaker size dependence, and is extrapolated to $T_{CG}= 0.145\pm 0.004$ (in units of $J$), while the spin crossing temperature exhibits a stronger size dependence extrapolated to $T_{SG}= 0.120\pm 0.006$. Hence, $T_{SG}$ is lower than $T_{CG}$ by about 15\%. For our $\xi/L$ data, we also tried a KT scaling with a logarithmic correction as was done in Ref.\cite{Campos06}. KT-like scaling can be ruled out, however, when our lower temperature $\xi_{CG}/L$ and $\xi_{SG}/L$ data, which were not available to Ref.\cite{Campos06}, are included. This is particularly clear for  $\xi_{CG}/L$ data where the curves are not ``merging''\cite{Campos06} nor ``marginal'' \cite{LeeYoung07}, but splay out below $T_{CG}$.

 The Binder ratios are shown in Fig.3 for the chirality (a), and for the spin (b). The chiral Binder ratio $g_{CG}$ exhibits a negative dip which deepens with increasing $L$. The data of different $L$ cross on the {\it negative\/} side of $g_{CG}$. These features indicate a finite-temperature transition in the chiral sector. By extrapolating the dip temperature $T_{dip}$ and the crossing temperature $T_{cross}$ for pairs of successive $L$ values to $L=\infty$, $T_{CG}$ is estimated to be $T_{CG}=0.145\pm 0.005$: See Fig.2. The estimated $T_{CG}$ is fully consistent with the one estimated above from $\xi_{CG}/L$. The peculiar form of $g_{CG}$ with a negative dip is consistent with the occurrence of a one-step-like replica-symmetry breaking as suggested by Hukushima and Kawamura \cite{HukuKawa00,HukuKawa05}. Indeed, the calculated chiral-overlap distribution below $T_{CG}$ (not shown here) develops a prominent central peak at $q_\chi =0$ similar to the one reported in Refs.\cite{HukuKawa00,HukuKawa05}.
\begin{figure}[ht]
\begin{center}
\includegraphics[scale=0.9]{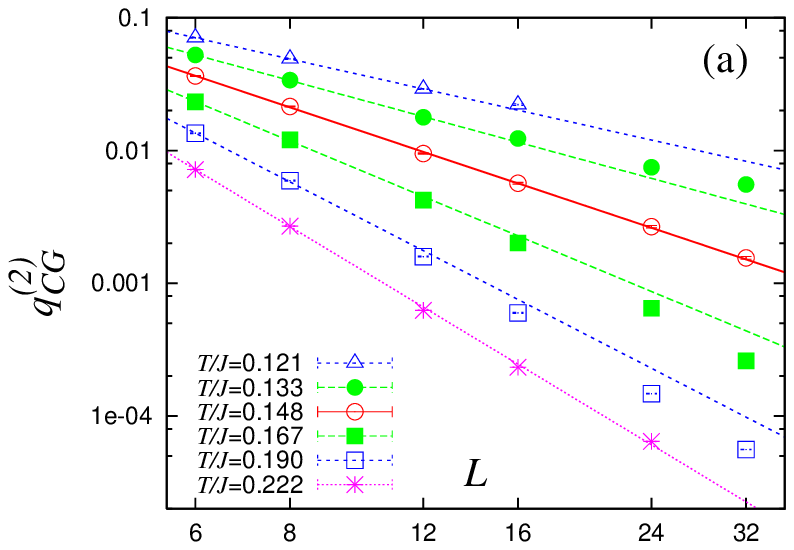}
\includegraphics[scale=0.9]{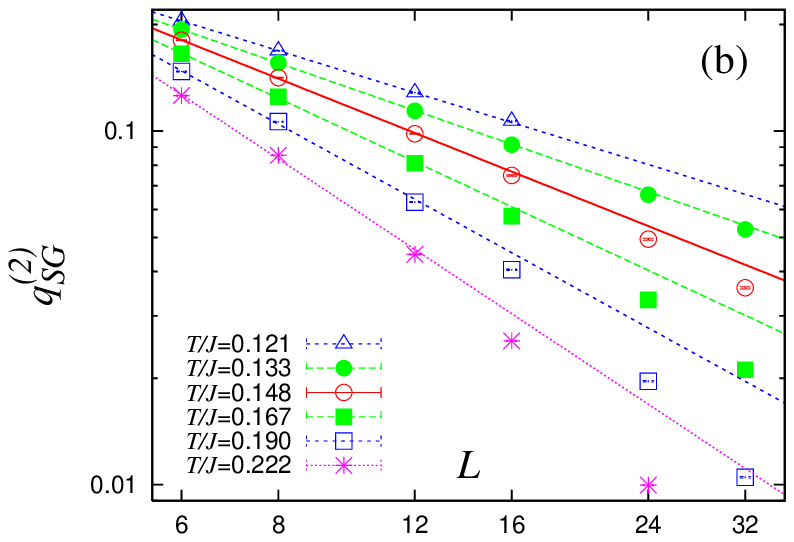}
\end{center}
\caption{Size dependence of the chiral-glass order parameter $q_{CG}^{(2)}$ (a),  and the spin-glass order parameter $q_{SG}^{(2)}$ (b). Straight lines in the figures are drawn by fitting the three data points of smaller sizes, $L=6,8$ and 12.
}
\end{figure}

 By contrast, the corresponding spin Binder ratio $g_{SG}$ does not exhibit crossing nor merging in the temperature range studied, suggesting that the SG transition temperature, if any, occurs below $T\simeq 0.12$. Meanwhile, as the size $L$ is increased, $g_{SG}$ develops more and more singular form at low temperature, indicating that the associated overlap distribution significantly changes its shape at low temperature. If one recalls the fact that $g_{SG}$ takes a value unity at $T=0$, $g_{SG}$ is expected to develop a negative dip at a lower $T$ (of $\lsim 0.12$) accompanied by an upturn toward $T=0$. 
This feature of $g_{SG}$ suggests the occurrence of a SG transition at a nonzero temperature, $T_{SG}\lsim 0.12$. 

 Pixley and Young recently criticized that the Binder ratio might not be an appropriate quantity in studying the ordering of vector SG, arguing that the large number of the order-parameter components ($3^2=9$ in the Heisenberg SG) might lead to a trivial Gaussian distribution even below $T_g$ \cite{PixleyYoung08}. In fact, mere the large number of order-parameter components does not lead to such a trivial behavior: For example, in a simple 3D $O(n)$ ferromagnet with $n=6$ \cite{Loisson} and 10 \cite{Viet}, the spin Binder ratio exhibits  at $T_c$ a clear crossing behavior characteristic of the standard long-range ordered phase, quite different from the one of Fig.3(b), presenting counter-examples to the criticism of Ref.\cite{PixleyYoung08}. 
Hence, the peculiar behavior of $g_{SG}$ observed here in Fig.3(b) should be regarded as a manifestation of essential features of the SG ordering.

 In Fig.4, we show the size dependence of the CG and SG order parameters on a log-log plot. 
As can be seen from Fig.(a), $q_{CG}^{(2)}$ exhibits a linear behavior at a temperature $T=0.148$, a downward curvature characteristic of a disordered phase at higher $T$, and an upward curvature characteristic of a long-rage ordered state at lower $T$. Thus, the data of $q_{CG}^{(2)}$ are consistent with a CG transition occurring at $T_{CG}= 0.148\pm 0.005$, consistently with the results of $\xi_{CG}/L$ and $g_{CG}$.

 The SG order parameter $q_{SG}^{(2)}$ exhibits a significantly different behavior, {\it i.e.\/}, it exhibits a downward curvature characteristic of a disordered state at $T=0.148\simeq T_{CG}$, or even at $T=0.133<T_{CG}$. At our lowest temperature $T/J=0.121$ where we could equilibrate only smaller lattices of $L\leq 16$, the data exhibit a near linear behavior up to $L=16$, although it is not clear whether this linear behavior extends to larger $L$. Thus, our data  of $q_{SG}^{(2)}(L)$ are consistent with a SG transition occurring at $T_{SG}\lsim 0.13$, whereas, from the present data of $q_{SG}^{(2)}$ only, we cannot rule out the possibility that $T_{SG}$ is significantly lower than this.


 All the physical quantities studied here, including the correlation-length ratio, the Binder ratio and the glass order parameter, consistently indicate that the 3D Heisenberg SG with Gaussian coupling exhibits successive CG and SG transitions at $T_{CG}=0.145\pm 0.004$ and at $T_{SG}\lsim 0.12$. The SG order sets in at a temperature at least about 15\% below the CG order, hence, the occurrence of the spin-chirality decoupling. One may feel that the relative distance between $T_{CG}$ and $T_{SG}$ is not so large, but, in fact, it is a sizable difference, much larger than the one observed in other systems exhibiting the spin-chirality decoupling, {\it e.g.\/}, the 2D regular frustrated {\it XY\/} model where the difference is known to be less than 1\% \cite{FFXY,2DXYSG}. While the SG order in the 3D Heisenberg SG occurs at nonzero temperature, as is consistent with the recent numerical works \cite{Matsubara00,LeeYoung03,Campos06,LeeYoung07}, it should be stressed that whether $T_{SG}$ is zero or nonzero is irrelevant to the chirality scenario of Refs.\cite{Kawamura92,Kawamura07} as long as the spin-chirality decoupling occurs, {\it i.e.\/}, $T_{SG} < T_{CG}$.  Our present result then supports the chirality scenario of SG transition.

 The authors are thankful to  I.A. Campbell and H. Yoshino for useful discussion. This study was supported by Grant-in-Aid for Scientific Research on Priority Areas ``Novel State of Matter Induced by Frustration'' (19052006). We thank ISSP, Tokyo University and YITP, Kyoto University for providing us with the CPU time.

\end{document}